\documentclass[aps,prl,twocolumn,letterpaper,showpacs,floatfix,groupedaddress,preprintnumbers,amsmath,amssymb]{revtex4}
\usepackage{graphicx}
\usepackage{dcolumn}% Align table columns on decimal point
\usepackage{bm}% bold math

\begin{document}

\title{Absorption and Emission in the non-Poisson case.}

\author{Gerardo Aquino$^1$, Luigi Palatella$^2$, Paolo Grigolini$^{1,2,3}$}
\address{$^{1}$Center for Nonlinear Science, University of North Texas,
   P.O. Box 311427, Denton, Texas 76203-1427 }
\address{$^{2}$Dipartimento di Fisica dell'Universit\`a di Pisa and
INFM, via Buonarroti 2, 56126 Pisa, Italy}
\address{$^{3}$
Istituto dei Processi Chimico Fisici del CNR
Area della Ricerca di Pisa, Via G. Moruzzi 1,56124 Pisa, Italy}
\date{\today}

\begin{abstract} 
This letter adresses the challenging problems posed to the
Kubo-Anderson (KA) theory by the discovery of intermittent resonant
fluorescence with a  non-exponential distribution of waiting
times. We show how to extend the KA theory from aged to aging systems, aging for a very extended time period or even forever, being a crucial consequence of non-Poisson statistics.
\end{abstract}

\maketitle

In the last few years, as a
consequence of an increasingly faster technological advance, it has
become clear that the conditions of ordinary statistical mechanics assumed by
the lineshape theory of Kubo and Anderson (KA) \cite{KA},  are violated  by some of the  new
materials. For instance,  the experimental research work of Neuhauser
\emph{et al.} \cite{bawendi} has established that the fluorescence emission of single nanocrystals exhibits interesting intermittent behavior, namely, a sequence of  ``light on'' and ``light off'' states, departing from Poisson statistics.  In fact, the waiting time distribution in both states is
non-exponential, and it shows a universal power law behavior
\cite{kuno}.  In this paper, for simplicity, we assign to both states
the same waiting time distribution 
\begin{equation}
\label{baby}
\psi(t) = (\nu -1) \frac{T^{\nu-1}}{(t+T)^{\nu}},
\end{equation}
with  $1 < \nu < \infty$. 
The parameter $T > 0$ is introduced for the purpose of making $\psi(t)$  
finite at $t = 0$ so as to ensure its normalization. We shall focus on the case when $\nu < 3$.
In accordance with Brokmann \emph{et al.} \cite{brokmann}, the experimental condition $\nu <2$ implies that the observed waiting time distribution depends on the time at which observation begins. Let us assume that the probability of the first jump from  the ``on'' (``off'') to the ``off'' (``on'') state is given by Eq. (\ref{baby}), if the observation begins at $t=0$. If  the observation begins at a later time $t_{a} > 0$, the distribution of the sojourn times, before the first jump, turns out to be different from Eq. (\ref{baby}): it is  $t_{a}$ dependent and, for this reason,  is denoted by $f(t,t_{a})$. This is the property responsible for the breakdown of the ordinary KA theory: it is the aging effect  on which we focus our attention in this letter. It is worth noticing that when $\nu > 2$, this aging effect is still present, in a less dramatic form, given the fact that  a stationary condition exists, even if the regression to it takes a virtually infinite time if $\nu < 3$ \cite{aging}.

The authors of Ref.  \cite{barkai} showed how to derive the absorption lineshape in the case $\nu  > 2$, when the stationary condition applies, and evaluated the form that the spectrum would have, immediately after switching on the radiation field, when the non-stationary condition $\nu < 2$  holds true.  Here we illustrate a way to evaluate the time evolution of the absorption spectrum, so as to take into account the aging effects of Brokmann \emph{ et al.}, with $\nu < 2$, as well as those of Ref. \cite{aging}, with $\nu >2$. We use the following stochastic equation:
\begin{equation}
\label{multiplicative}
\frac{d}{dt} \mu(t) = i(\omega_{0} + \xi(t))\mu(t).
\end{equation}
The quantity $\mu(t)$ is a complex number, corresponding to the
operator $|e\rangle\langle g|$ of the more rigorous quantum mechanical treatment
\cite{advchemphys}, $|e\rangle$ and $|g\rangle$ being the excited and the groundstate, respectively, $\omega_{0}$ is the energy difference between the
excited and the ground state, and $\xi(t)$ denotes the energy
fluctuations caused by the cooperative environment of this system.  In
the presence of the coherent excitation, Eq. (\ref{multiplicative}) becomes
\begin{equation}
\label{radiation}
\frac{d}{dt} \mu(t) = i(\omega_{0} + \xi(t))\mu(t) + k exp(i \omega t),
\end{equation}
where $\omega $ denotes the radiation field frequency.  It is
convenient to adopt the rotating-wave approximation. Let us express
Eq. (\ref{radiation}) by means of the transformation $\tilde \mu(t) =
exp(i \omega t) \mu(t)$. After some algebra, we get a simple equation
of motion for $\tilde \mu(t)$.  For simplicity we denote $\tilde
\mu(t)$ with the symbol $\mu(t)$ again, thereby making the resulting equation read:
\begin{equation}\label{interactionpicture}
\frac{d}{dt} \mu(t) = i(\delta + \xi(t))\mu(t) + k ,
\end{equation}
where   $\delta = \omega_{0} - \omega$.
%\begin{equation}
%\delta = \omega_{0} - \omega.
%\end{equation}
 The reader can easily establish the connection between this picture
and the stochastic Bloch equation of Ref. \cite{advchemphys} by
setting $\mu = v + i u$. Note that the three components of the Bloch vector in Ref. \cite{advchemphys}, $(u,v,w)$, are related to the rotating-wave representation of the 
density matrix $\rho$, $v$ and $u$ being the
imaginary and the real part of $ e^{-i\omega t}\rho_{ge}$, and $w$ being
defined by $w \equiv (\rho_{ee} - \rho_{gg})/2$.  Note that the
equivalence with the picture of Ref. \cite{advchemphys} is established
by assuming the radiative lifetime of the excited state to be infinitely large
and the Rabi frequency $\Omega \equiv k$ vanishingly small.

  It is straightforward to integrate Eq. (\ref{interactionpicture}), thereby getting 
\begin{equation}
\label{directintegration}
 \mu(t)  = k \int_{0}^{t} dt' e^{i \int_{t'}^{t}\xi(t'')dt''} e^{i\delta(t-t')},
\end{equation}
with the dipole $\mu = 0$, when the exciting radiation is turned on. 
Now we have to address the
intriguing issue of averaging Eq. (\ref{directintegration}) over a set
of identical systems, in such a way as to take aging effects into account \cite{brokmann,barkai2,aging}.  In fact the averaging process
 turns Eq. (\ref{directintegration}) into
\begin{equation}
\label{averaging}
\langle \mu(t)\rangle = k \int_{0}^{t} dt' \langle e^{i \int_{t'}^{t}\xi(t'')dt''} \rangle_{t'}e^{i\delta(t-t')},
\end{equation}
with the subscript $t'$ denoting that the system, brand new at $t=0$, is $t'$-old when we evaluate the corresponding characteristic function, thereby implying that the distribution of waiting times before the first jump, is not $\psi(t)$, and   $f(t,t')$ has to be used instead. Our numerical approach rests on the following prescriptions. Using, the distribution of Eq. (\ref{baby}), we
run $N$ distinct sequences $\{\tau_{i}\}$, with $N \gg 1$. For any of these $N$ sequences the sojourn in one of the two states begins exactly
a time $t= 0$ and ends at time $t = \tau_{1}$. For $N/2$ of these sequences we use the ``light on" as initial condition, and for $N/2$ the  ``light off" state. Let us consider the former type of trajectories for illustration purpose.  The  ``light on" state begins at $t= 0$ and ends at $t = \tau_{1}$, at which time  the ``light off" state
begins, ending at time $t = \tau_{1} + \tau_{2}$, and so on.   With
this numerical method we produce a set of fluctuations $\xi(t)$. Then,
we create a set of diffusion trajectories $\int_{t'}^{t} \xi(t'')dt''$, and hence
the set of exponentials $e^{  i \int_{t '}^{t}\xi(t'')dt''}$. Since all trajectories of this set are $t'$-old, the resulting numerical average is automatically equivalent to evaluating $\langle e^{  i \int_{t'}^{t}\xi(t'')dt''}\rangle_{t'}$ with the waiting time distribution $f(t,t^{\prime})$. We point out that the number of photons emitted at time $t$ is determined by $N(t)= \langle \mu(t)\mu^{*}(t)\rangle$. It is straightforward to prove that the rate of photons emitted, namely, $R(t) \equiv dN/dt$, obeys the relation
\begin{equation}
R(t) = 2k Re \langle \mu(t)\rangle .
\end{equation}
Thus, we conclude that the real part of $\mu(t)$ can be used to denote
either emission or absorption at time $t$.
We note that the approximation ensuring the equivalence between our 
picture and Ref. \cite{advchemphys}, for large photon count, has also the effect of making the absorption identical to the emission spectrum.
\begin{figure}[!h]
\includegraphics[width=7.8 cm]{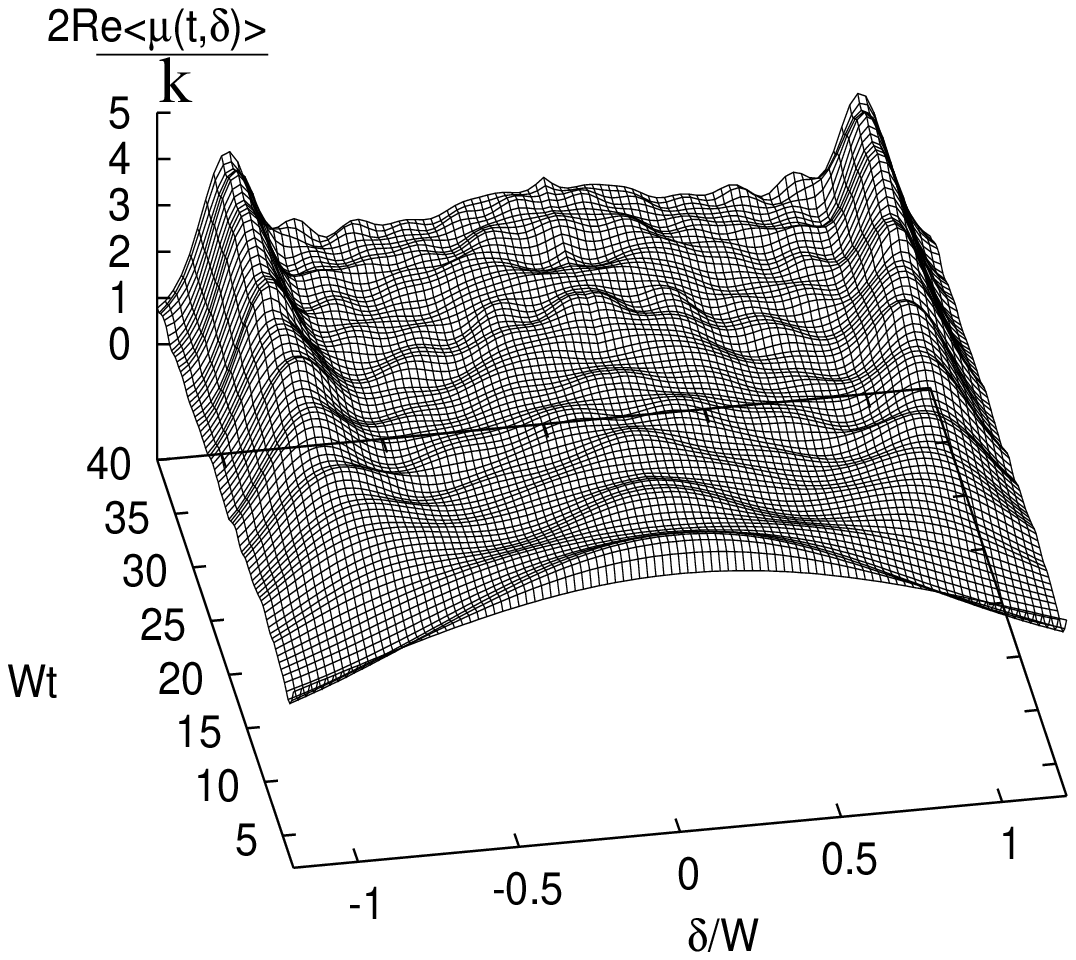}
\includegraphics[width=7.8 cm]{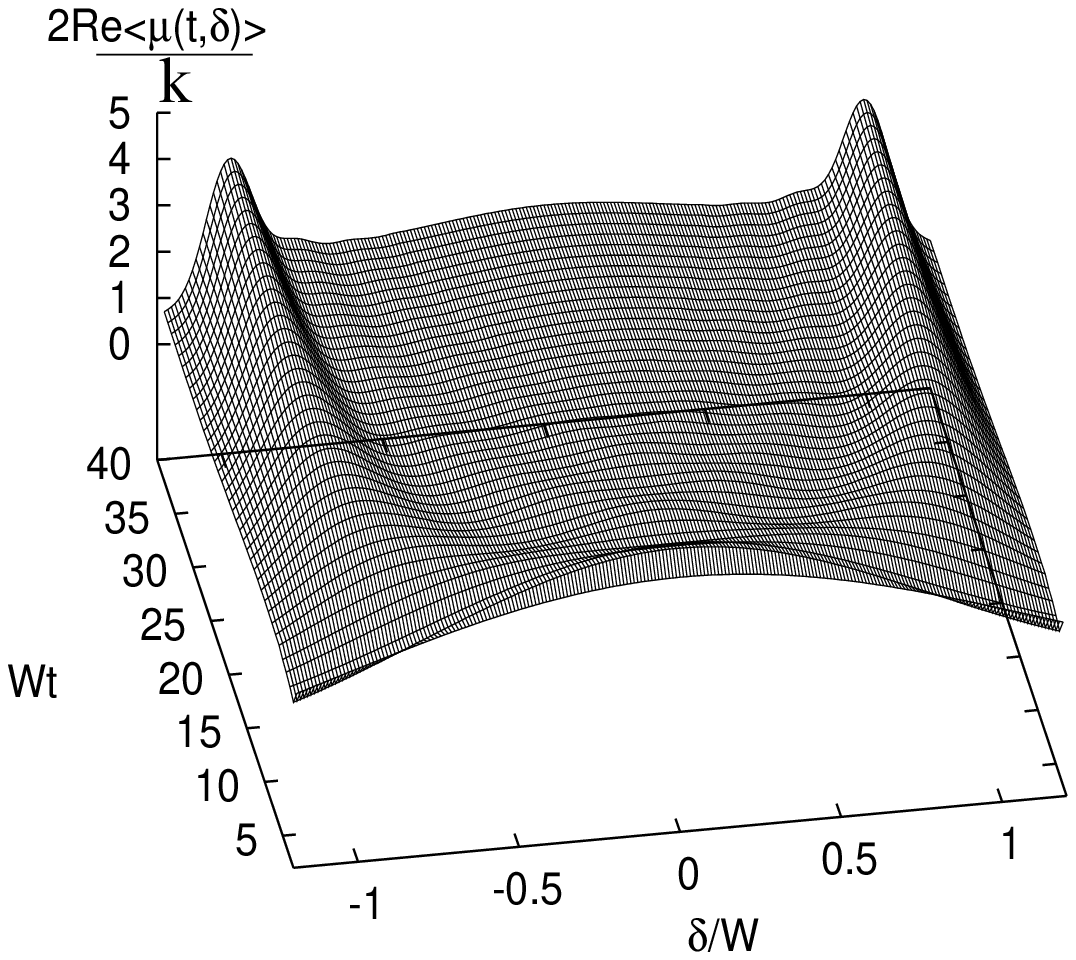}
%\vspace{-0.1 cm}
\caption{\label{fig1}  The numerical (top) versus the theoretical (bottom) 
absorption spectrum with  $\nu = 2.5$,  $T \cdot W=0.4$. We assign to  the fluctuation $\xi(t)$ the intensity $W=1$.}
%\vspace{-0.1 cm}
\end{figure}
\begin{figure}[!h]
\includegraphics[width=7.8 cm]{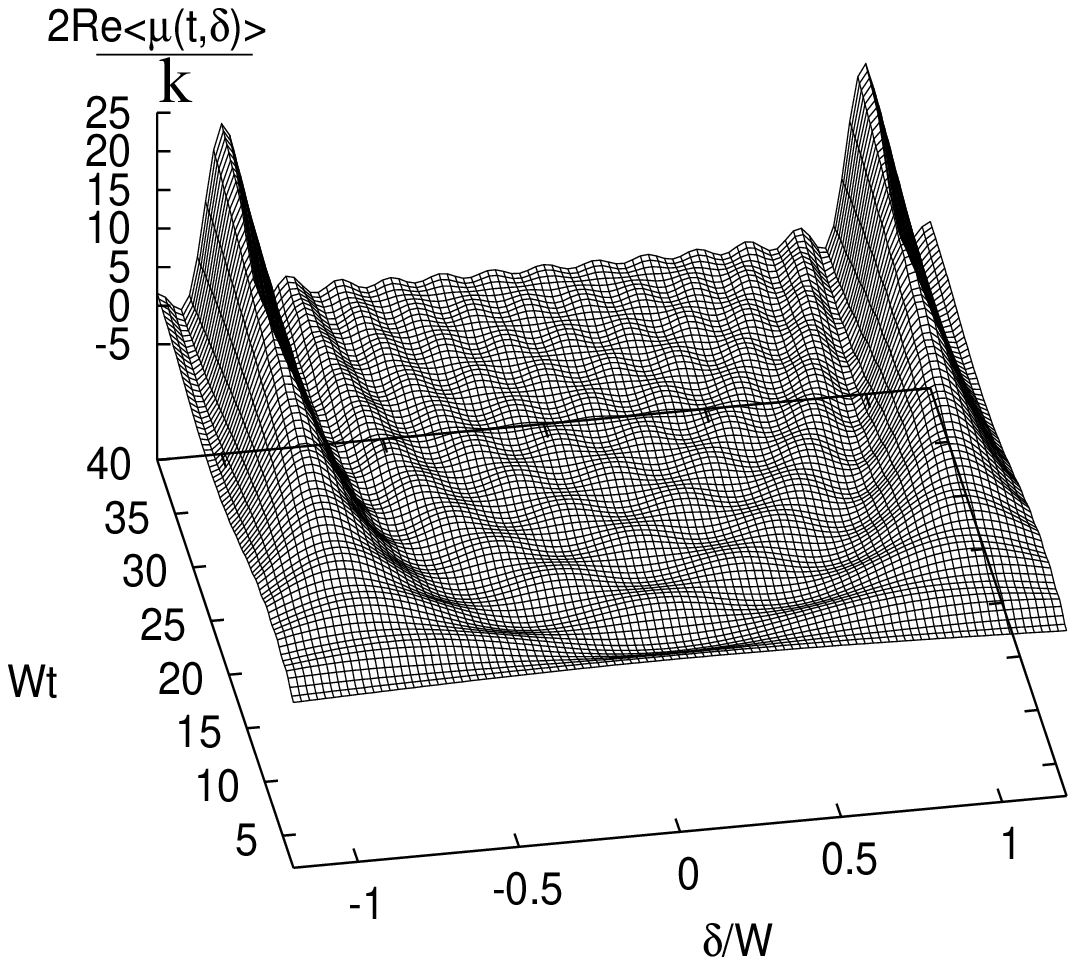}
\includegraphics[width=7.8 cm]{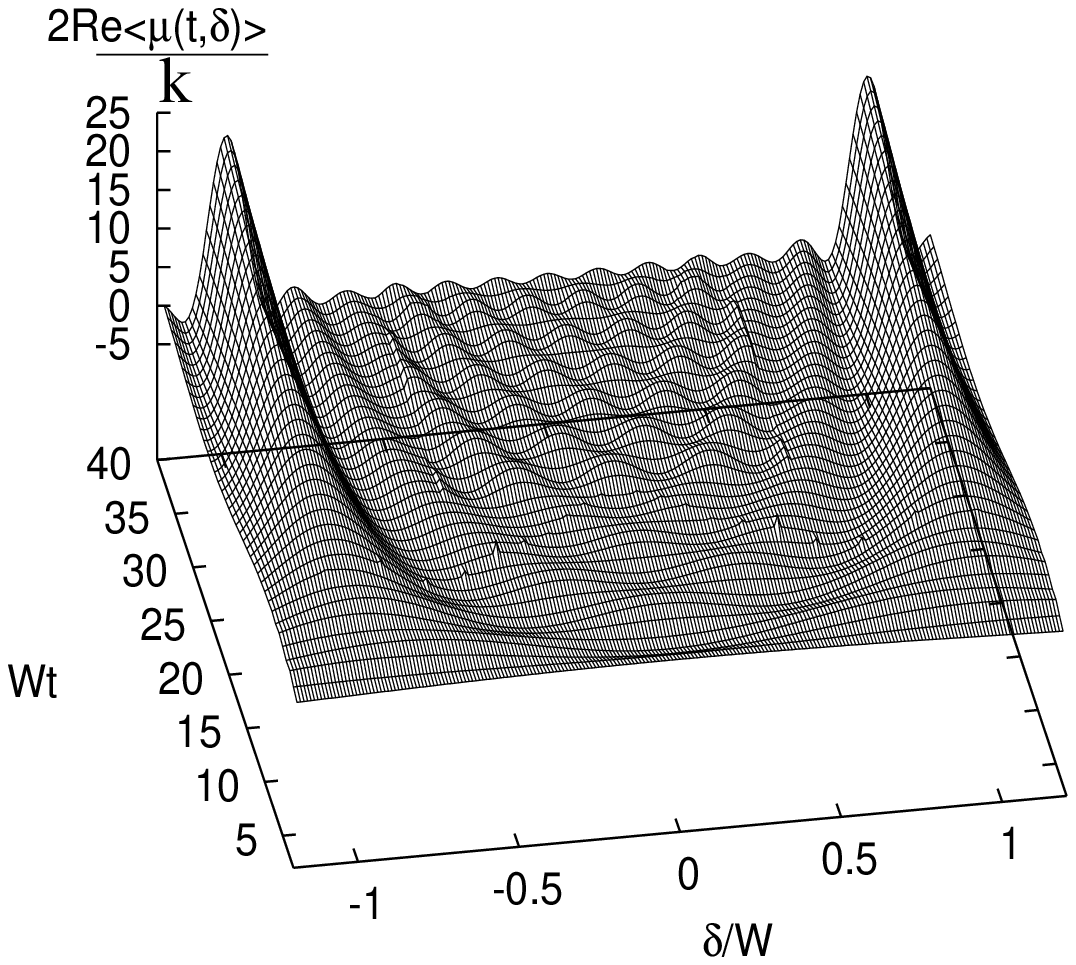}
%\vspace{-0.0 cm}
\caption{\label{fig2} The numerical (top) versus the theoretical (bottom) 
absorption spectrum with  $\nu = 1.5$,  $T \cdot W=0.4$. We assign to  the fluctuation $\xi(t)$ the intensity $W=1$.}
%\vspace{-0.0 cm}
\end{figure}
\begin{figure}[!h]
\includegraphics[width=5.8 cm,height=7.6 cm, angle=270]{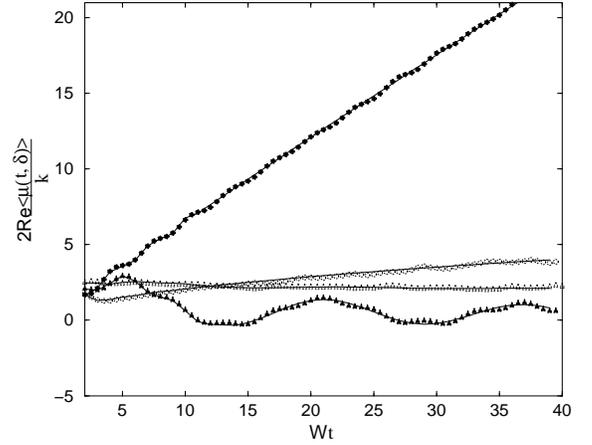}
%\vspace{-0.0 cm}
\caption{\label{fig3} Comparison between numerical  and theoretical  absorption spectra for different values of the de-tuning parameter $\delta$ and of the power index $\nu$. Moving from the top to the bottom in the
right hand portion of the figure,  $\delta = 1$, $\nu = 1.5$, $\delta = 1$, $\nu = 2.5$, $\delta = 0.6$, $\nu = 2.5$, $\delta = 0.6$, $\nu = 1.5$. The numerical curves are denoted by black circles, open circles, open triangles and black triangles, respectively, and in this scale are almost indistinguishable from the full lines denoting the corresponding theoretical results. }
%\vspace{-0.0 cm}
\end{figure}
Fig. \ref{fig1}  illustrates the rate of absorption changing upon change of
time, when $\nu = 2.5$. We see that the spectrum changes from a
Lorentzian shape centered at $\omega = \omega_{0}$ to a bi-modal shape
at later time, in accordance with earlier theoretical work \cite{paolone,earlierwork}. The authors of Ref. \cite{paolone} studied the aged condition $f(t,\infty)$ with $2 < \nu < 3$, and found that the corresponding spectrum has two sharp peaks. Klafter and Zumofen proved that  the same case with $f(t, t_{a} = 0) = \psi(t) $ yields L\'{e}vy diffusion, thereby implying an exponential characteristic function, and consequently a Lorentzian spectrum. It is remarkable that our theoretical  perspective establishes a connection between  the prediction of Zumofen and Klafter, valid at short times,  and that of  Ref. \cite{paolone}, valid at large times.  Remarkably, we  evaluate numerically also the spectrum time evolution in the case $\nu < 2$, where the process of aging keeps going on forever without ever reaching the stationary condition, which does not exist, in this case (see Fig. \ref{fig2}). 

Let us show now how to reproduce these numerical results with a proper theoretical treatment. Let us consider as an example a  single trajectory starting a time $t^{\prime}$ with $\xi=+W$ and ending at time $t$ with the same positive value after  $n$ switches of the variable $\xi$ between its two values $\pm W$. In this case the integrand of (\ref{directintegration}), has the following form:
\begin{equation}\label{example}
e^{(\alpha_+(t_1-t^{\prime})+\alpha_-(t_2-t_1)+.....+\alpha_+(t-t_{n}))}\end{equation}
where $\alpha_{\pm}=i(\delta\pm W)$ and $n$ can only be even.

 To determine the contribution to $\langle \mu(t)\rangle$ of Eq. (\ref{averaging}) stemming from this kind of trajectories, we have to average the term  (\ref{example}) on the set of  all   possible sequences
of  the variable $\xi(t)$ running from  $t^{\prime}$ to  $t$ with $n$ switches, to sum over $n$ and then to carry out the integration on $t^{\prime}$.\\
 To this purpose,  as earlier stated, besides the probability density $\psi(\tau)$ of having an interval $\tau$ between two generic consecutive switching events, it is necessary to use $f(\tau,t^{\prime})$, this being  the conditional probability density  that, fixed a time $t^{\prime}$, the first next switching event of the variable $\xi(t)$ occurs at time $t^{\prime}+\tau$.
  The result of this procedure turns the contribution to (\ref{averaging}) 
 into:
\begin{equation}\label{convol}
\begin{split}
  &\frac{k}{2}\int_0^t dt^{\prime}\left( \rule{0 cm}{0.65 cm}  F(t-t^{\prime},t^{\prime})e^{\alpha_+(t-t^{\prime})}   + \sum_{n=1}^{\infty} \int_{t^{\prime}}^t dt_1  f(t_1-t^{\prime},t^{\prime}) \right.\\
 &\left. \cdot \;  e^{\alpha_{+}(t_1-t^{\prime})} \int_{t_1}^t dt_2\psi(t_2-t_1)e^{\alpha_{-}(t_2-t_1)}\int_{t_2}^t dt_3  \psi(t_3-t_2) \right.\\
%&\left.  \cdot \int_{\tau_2}^td\tau_3 \psi(\tau_3-\tau_2) e^{\alpha_+(\tau_3-\tau_2)}\right.\\
 &\left. \cdot \; e^{\alpha_{+}(t_3-t_2)}\cdots  \int_{t_{2n-1}}^t dt_{2n}\psi(t_{2n}-t_{2n-1})e^{\alpha_-(t_{2n}-t_{2n-1})} \right.\\
& \left. \hspace{0.5 cm} \cdot \, \, \Psi(t-t_{2n}) e^{\alpha_+(t-t_{2n})} \rule{0 cm}{0.65 cm} \right),\end{split}
\end{equation}
where, in addition to the crucial probability density $f(\tau,t')$, we have used\begin{equation}
\label{Psi} \Psi(\tau)=1-\int_0^{\tau}\psi(\tau')d\tau',
\end{equation}
which is the conventional probability that no switch occurs 
for a generic interval of time $\tau$, and
\begin{equation}\label{F}
 F(\tau,t^{\prime})=1-\int_0^\tau f(\tau',t^{\prime})d\tau',
\end{equation}
which is the corresponding aging property, depending on $ f(\tau,t')$, 
and indicating therefore the conditional  probability that, fixed $t^{\prime}$, no switch occurs between $t= t^{\prime}$ and $t = t^{\prime}+ \tau$.
 The overall factor of $1/2$  of the contribution (\ref{convol}) is a consequence 
of the fact that at time $t^{\prime}$ the fluctuation $\xi(t)$, supposed to be 
positive,  can get with the same probability the negative value.
Let us address now the problem of finding an exact analytical expression for the crucial property $f(\tau,t')$. 
The exact expression for
$f(\tau,t^{\prime})$, is
\begin{equation}
f(\tau,t^{\prime})=\int_{0}^{t^{\prime}}d\tau^{\prime} G(t' - \tau^{\prime})\psi (\tau+ \tau^{\prime}),  \label{exact}
\end{equation}
where 
\begin{eqnarray}\label{g}
&&G(t)\equiv \delta(t)+\psi(t)+\sum_{n=2}^{\infty }\int_{0}^{t}dt_{1}\psi (t_{1})\\
\nonumber && \cdot \int_{t_{1}}^{t}dt_{2}\psi (t_{2}-t_{1})...\int_{t_{n-2}}^{t}dt_{n-1}\psi (t-t_{n-1}).  
\end{eqnarray}
 It is straightforward to find the Laplace transform of $G(t)$. This is given
by 
\begin{equation}
\hat{G}(u)=\sum_{n=0}^{\infty}\hat{\psi}(u)^{n}=\frac{1}{1-\hat{\psi}(u)},
\label{laplacetransformofg}
\end{equation}
where $\hat \psi(u)$ denotes the Laplace transform of $\psi(t)$. Thus, the Laplace transform of (\ref{exact}) with respect to $t^{\prime}$ reads: 
\begin{equation} \label{exactlaplace}
\hat{f}(\tau, u^{\prime})=\frac{1}{1-\hat \psi (u^{\prime})}e^{u^{\prime}\tau}\left[ \hat \psi
(u^{\prime})-\int_{0}^{\tau}e^{-u^{\prime} y}\psi (y)dy\right],  
\end{equation}
thereby yielding:
 \begin{equation}
\label{fF}
 f(\tau,t^{\prime})={\cal L}^{-1}[\hat{f} (\tau,u^{\prime})].
\end{equation}
With this expression the prescription necessary to evaluate the contribution to $\langle\mu(t)\rangle$ of Eq. (\ref{averaging}) of the  trajectories beginning in the ``light on" state at $t'$ and ending in the same state at $t$, is completed.

As a last step,  we calculate the Laplace transform of (\ref{convol}), and of the
equivalent expressions  for all the other possible conditions of motion from  $t^{\prime}$ to $t$, namely, with the noise $\xi(t)$ moving  from $+W$ to $-W$, from $-W$ to $+W$ and, finally, from $-W$ to $-W$, as well. This procedure yields, as a  final result, the Laplace transform of $R(t)$, denoted by $\hat R(u)$, which is proved to have the following analytical expression:
\begin{equation}\label{analytic}
\hat R(u) = k {\cal L}[\langle \mu(t)+\mu^{*}(t)\rangle]=\frac{k^2}{2} (   A_+(u)  +  A_-(u)  +  C(u)).
% \frac{k^2}{2} (  \hat A(u)  + \hat B(u)  + \hat C(u)).
\end{equation}
The explicit expression for $A_+(u)$, which is calculated taking into account the contribution to  $\langle\mu(t)\rangle$ of  those  trajectories ending at   time $t$ with a  positive value  $\xi(t)=+W$ for the flucuating variable, is
\begin{equation}
A_+(u) \equiv \frac{\hat \Psi(u-\alpha_{+})[\hat f_{+}(u) \hat \psi(u-\alpha_{-})+\hat f_{-} (u) ] }{1 - \hat \psi(u-\alpha_{+})
\hat \psi(u-\alpha_{-})} + \hat F_{+}(u) ,
\end{equation}
with:
\begin{equation}
 \hat{\Psi}(u-\alpha_{\pm})={\cal L}\;[\Psi(t)e^{\alpha_{\pm}t}]=\frac{1-\hat{\psi}(u-\alpha_{\pm})}{u-\alpha_{\pm}}, 
\end{equation}
\begin{eqnarray}
\hat f_{\pm}(u) & \equiv & {\cal L}\;[\int_0^t dt^{\prime}f(t-t^{\prime},t^{\prime})e^{\alpha_{\pm}(t-t^{\prime})}]\\
\nonumber &=&  \frac{\hat \psi(u)  - \hat \psi(u - \alpha_{\pm})}{-\alpha_{\pm}(1 - \hat{\psi}(u))}
\end{eqnarray}
%\vspace{0.1 cm}
and\begin{equation}
%\begin{eqnarray}
\hat F_{\pm}(u) \equiv {\cal L}\;[\int_0^t dt^{\prime}F(t-t^{\prime},t^{\prime})e^{\alpha_{\pm}(t-t^{\prime})}]=\frac{1/u - \hat f_{\pm}(u) }{(u - \alpha_{\pm})}.
\end{equation}
$A_-(u)$ takes into account the contribution of all  trajectories ending   at time $t$ with the  negative value   $\xi(t)=-W$, for the fluctuating variable and is  derived from the expression for $A_+(u)$ by replacing everywhere $\alpha_{\pm}$ with $\alpha_{\mp}$ (sending  $\hat{f_{\pm}},\hat{F_{\pm}} \to \hat{f_{\mp}},\hat{F_{\mp}}$).
Thus, $k(A_+(u)+  A_-(u))/2$ represents the laplace transform of $\langle\mu(t)\rangle$. 
Finally $k C(u)/2$ is the Laplace transform of  $\langle \mu^{*}(t)\rangle$, and it is derived from the earlier expression for $k(A_+(u)+  A_-(u))/2$ by replacing everywhere $\alpha_{\pm}$ with $-\alpha_{\pm}$.
 To establish a
comparison with the result of the numerical experiment we have
anti-Laplace transformed the analytical expression of
Eq. (\ref{analytic}) using a Talbot  algorithm implemented on Mathematica 5.0. The result of this procedure is illustrated by both
Fig. \ref{fig1} and Fig. \ref{fig2}, where the analytical predictions are compared  to  the corresponding  numerical experiments. The overall qualitative agreement is remarkably good. 
Apparently, in Fig. \ref{fig1} the departure of the theoretical calculation 
from the fine structure of the numerical result is larger than in 
Fig. \ref{fig2}.
 Actually, the numerical algorithm  produces fluctuations of the same
intensity in both cases, although the more detailed scale of Fig. \ref{fig1} makes
them appear larger.
 To establish beyond any doubt also the quantitative  agreement, in Fig. \ref{fig3} we compare the two predictions for specific values of the de-tuning parameter $\delta$. This shows that also the quantitative agreement is very satisfactory. Furthermore, when a  stationary condition exists, it is 
straightforward to prove that
Eq. (\ref{analytic}) recovers, as asymptotic-time limit,
the same result as that of \cite{barkai}.
\\ \indent
   In summary, we could afford a complete picture of the aging process, 
going beyond either the short-time or asymptotic-time limitations of Ref. \cite{barkai}. This letter 
shows also that these aging spectra are a consequence of the fact 
that the field acting at time $t^{\prime}$ makes the photon emission depend on 
$t^{\prime}$-old trajectories. This yields
Eq. (\ref{averaging}), with the subscript $t'$ working as an age indicator.
This formula intertwines the system to the radiation field, thereby 
violating the ordinary linear response prescription, which is 
recovered in fact in the Poisson case.\\
\indent  GA and PG thankfully acnowledge ARO  for financial support through Grant DAAD19-02-1-0037.

\end{document}